\documentclass[%
preprint,
aip,
amsmath,amssymb,
aps,
]{revtex4-2}
\usepackage{dcolumn}
\usepackage{bm}
\usepackage{graphicx}

\newcommand{\jgr}{J. Geophys. Res.}
\newcommand{\apjl}{Astrophys. J. Lett.}

\newcommand{\araa}{Ann. Rev. Astron. Astrophys.}
\newcommand{\mnras}{Mon. R. Astron. Soc.}
\newcommand{\ssr}{Space Sci. Rev.}

\usepackage{color}

\newcommand{\crd}{\color{black}}
\newcommand{\cbk}{\color{black}}

\begin{document}
\title{A Hard Energy Spectrum in 3D Guide-Field Magnetic Reconnection} 
\author{Masahiro Hoshino}
\affiliation{Department of Earth and Planetary Science, The University of Tokyo, Tokyo 113-0033, Japan.}
\email{hoshino@eps.s.u-tokyo.ac.jp}
                       
\begin{abstract}\
Magnetic reconnection has long been known to be the most important mechanism not only for mixing the plasmas by changing the magnetic field topology but also for releasing the magnetic field energy into the plasma kinetic energy.  During magnetic energy release, it is possible for some of the heated plasma to be accelerated to energies much higher than the thermal energy.  Recently, the energy partitioning of the thermal and the nonthermal energy has been studied by using particle-in-cell (PIC) simulations, and it has been shown that the acceleration efficiency of nonthermal particles increases with increasing the plasma temperature, and the nonthermal energy density occupies more than $90 \%$ in the total heated plasma when the Alfv\'{e}n velocity is close to the speed of light $c$. However, the acceleration efficiency decreases as the guide magnetic field increases. So far the acceleration efficiency has been mainly studied in two-dimensional systems, but it is interesting to study three-dimensional effects where the patchy and turbulent reconnection can dynamically occur.  This study explores the effects of three-dimensional relativistic reconnection on a pair plasma with the guide magnetic field, utilizing three-dimensional (3D) PIC simulations.  The results indicate that the decrease in nonthermal particle production is smaller in 3D guide-field reconnection compared to 2D. More importantly, contrary to general expectation, 3D reconnection is capable of maintaining a hard nonthermal energy spectrum even in the presence of a strong guide magnetic field. 
\end{abstract}

\maketitle
\section{Introduction}
Magnetic reconnection is widely recognized as a crucial mechanism for plasma heating and particle acceleration in various explosive plasma phenomena in the Universe \cite{Blackman94,Birn07,Zweibel09,Hoshino12b,Uzdensky16}.
The magnetic field energy stored in the plasma sheet is released, which results in the thermalization of the plasma up to the equivalent temperature of the Alfv\'{e}n velocity. Furthermore, a portion of the plasma is accelerated to higher energies above the thermal temperature, forming a nonthermal component that can be approximated by a power-law spectrum.  

Since there is much observation evidence that the nonthermal particles associated with the hot thermalized plasma formed during the solar flares \cite{Lin03,Oka18} and the magnetospheric substroms in the Earth's magnetosphere \cite{Baker77,Hoshino01,Oieroset02}, it is believed that the generation of nonthermal particles in reconnection is a ubiquitous process in the \crd collisionless \cbk plasma environment.  Then the high-energy phenomena in a relativistic plasma medium observed around black holes, pulsar nebulae, and magnetars, etc., are also thought to be related to magnetic reconnection\cite{Lyutikov03,Kirk04,Madejski16,Blandford17}. Indeed, the numerical simulations have shown that the efficiency of the nonthermal particle acceleration is extremely high in such a relativistic plasma regime with $V_A \sim c$, and the formation of a hard energy spectrum has been discussed \cite{Zenitani01,Zenitani05b,Zenitani08,Jaroschek09,Liu11,Sironi11,Cerutti12,Bessho12,Cerutti13,Cerutti14,Sironi14,Guo14,Rowan17,Werner18,Ball18,Totorica20,French23}. 

While many observational and theoretical/computational studies have been conducted to study nonthermal particle acceleration in the plasma sheet, it was not necessarily investigated by focusing on the energy partitioning between the thermal energy density and the nonthermal energy density \cite{Eastwood13,Hoshino18}. Recently, we discussed the energy partitioning during magnetic reconnection with a pair plasma by using a two-dimensional particle-in-cell (PIC) simulation \cite{Hoshino22,Hoshino23}. It is found that the production of nonthermal particles increases with increasing Alfv\'{e}n speed for magnetic reconnection \cite{Hoshino22}. In the relativistic reconnection, where the Alfv\'{e}n speed is close to the speed of light, it has been shown that the nonthermal energy density can occupy more than $90 \%$ of the total kinetic plasma energy density for an anti-parallel magnetic field geometry without a guide magnetic field. In addition,we have discussed that the production of nonthermal particles decreases with the increase of a guide magnetic field \cite{Hoshino23}.

So far our study on the energy partitioning of thermal and nonthermal plasma has been limited to two-dimensional (2D) system \cite{Hoshino23} (Paper I), but it is important to investigate three-dimensional (3D) reconnection.  In 2D reconnection, the magnetic field ``reconnection'' can only occur in the neutral sheet, but 3D reconnection can occur not only in the neutral sheet but also outside the neutral sheet \crd\cite{Galeev86,Daughton11,Dahlin17,Guo21}\cbk. It has been shown that the three-dimensional reconnection has more dynamic time evolution, leading to patchy and turbulent reconnection \crd\citep{Daughton11,Dahlin17,Guo21}\cbk.
\crd
In addition, it has been discussed that the efficiency of electron acceleration in a 3D and non-relativistic reconnection is enhanced when a guide magnetic field is comparable to the reconnecting magnetic field component \citep{Dahlin17}. 
After these previous studies on the importance of 3D turbulent reconnection and plasma transport with a guide magnetic field, it is intriguing to investigate the nonthermal particle acceleration in a relativistic plasma with a more stronger guide magnetic field, because the patchy and turbulent reconnection is expected to be present in a wide range of plasma parameters in the plasma universe.
\cbk
Here we study the efficiency of the nonthermal particle acceleration of the 3D \crd relativistic \cbk reconnection as a function of the guide magnetic field by using 3D PIC simulations.  We show that the efficiency of the magnetic energy release in 3D is almost the same as that in 2D for a weak guide magnetic field, and that the efficiency of both 2D and 3D reconnection decreases with increasing the guide magnetic field.  The decrease in efficiency of 3D, however, is less than that of 2D.  More importantly, while the kappa index (the power law index) increases with increasing the guide magnetic field for 2D reconnection, the nonthermal spectra in 3D reconnection with a strong guide magnetic field regime retain a hard spectrum.

\section{Turbulent 3D Guide-Field Reconnection}
\begin{figure*}
\begin{center}
\includegraphics[width=16cm]{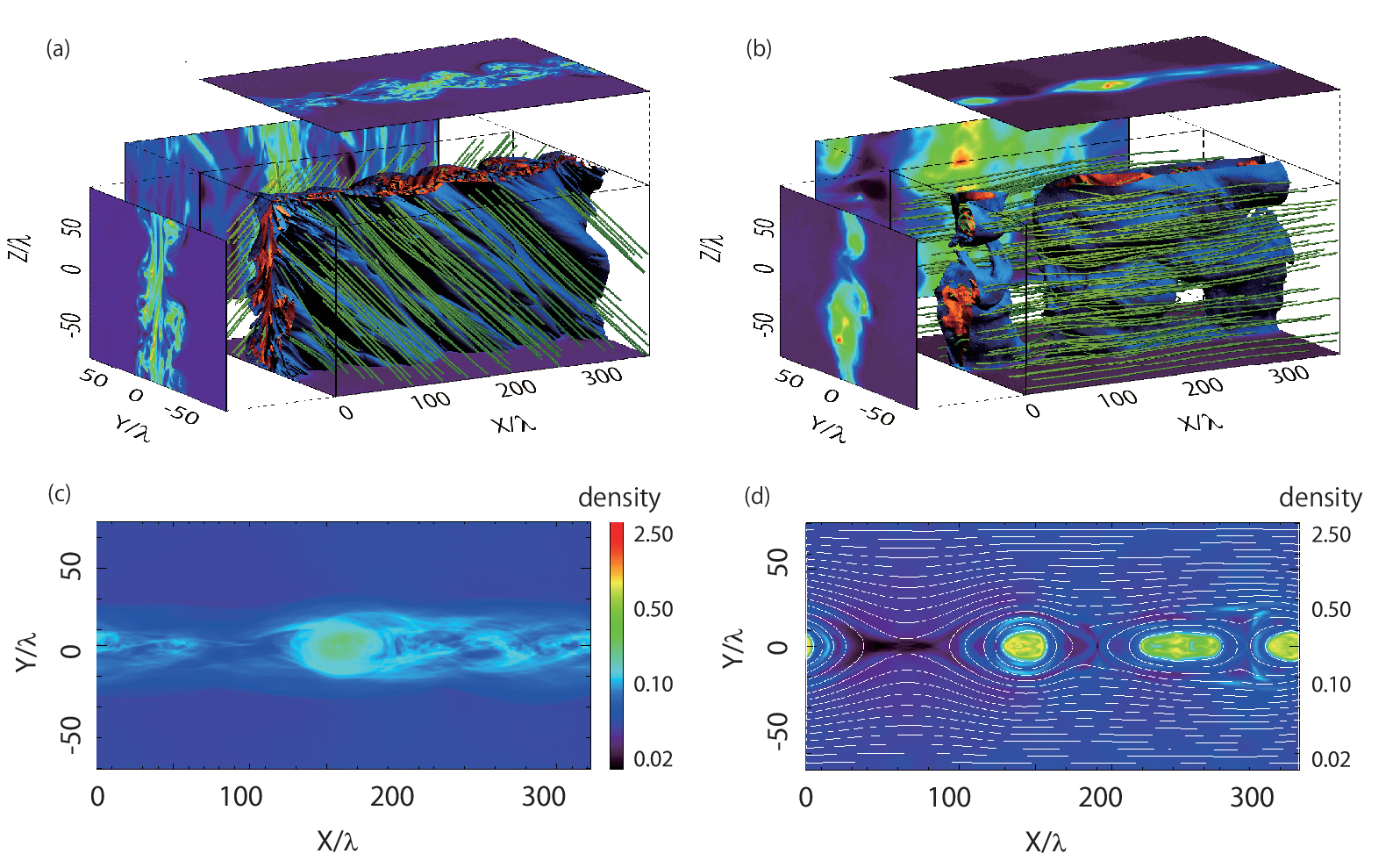}
\caption{Structure of magnetic reconnection with and without guide magnetic field. (a) 3D reconnection structure with the guide magnetic field $B_G/B_0=1$ at $t(V_A/\lambda)=366$. The red and blue isosurfaces show high and low plasma densities, respectively. The green lines are magnetic field lines, and two-dimensional contours in xy-, yz- and xz-plane are the slice at $z/\lambda$=80.7, $x$=0, and the neutral sheet $y=0$, respectively. (b) 3D reconnection structure without the guide magnetic field. (c) The integrated plasma density along the z-axis of case (a). (d) 2D reconnection with the guide magnetic field $B_G/B_0=1$ with the same parameter as case (a), and the white lines are the magnetic field lines.}
\label{fig:structure}
\end{center}
\end{figure*}

Our objective is to understand particle acceleration in three-dimensional reconnection with a guide magnetic field, but before discussing the particle acceleration, let us study the difference and similarity between 2D and 3D reconnection.  
We use basically the same 3D simulation code as that used in the study of the magnetorotational instability \cite{Hoshino15}. The numerical scheme with a semi-implicit time integration is the same as that used in our previous study of 2D reconnection \cite{Hoshino23}.  
We assume a periodic boundary in the $x$ and $z$ directions with the system size of $0 < x/\lambda < 322.5$ and $-80.7 < z/\lambda < 80.7$.  We use the so-called conducting wall in the $y$ direction with the upper and lower boundaries situated at $y= \pm 80.7 \lambda$, where $\lambda$ is the thickness of the initial plasma sheet. The computational grid size is $(N_x, N_y, N_z) = 4032 \times 2016 \times 2016$. The total number of particles is set as $N_p=1.6 \times 10^{11}$, and the average number of particles per grid size at the neutral sheet is $81$.  For the two-dimensional simulation, we use the same system size in the $x$ and $y$ directions as for the three-dimensional simulation, and the average number of particles per grid size at the neutral sheet is $972$.  These values are used to accurately calculate the magnetic energy dissipation process in the plasma sheet.

We adopt the Harris solution \citep{Harris62} for the pair plasma with the same mass $m$ and uniform temperature $T_0$ in space and focused on the energy partitioning of an idealized magnetic reconnection in a collisionless plasma system. 
The magnetic field ${\bf B}(y)$ and plasma density $N(y)$ are expressed as
\begin{equation}
  {\bf B}(y) = B_0 \tanh( y/\lambda) {\bf e_x} + B_G {\bf e_z},
\end{equation}
and
\begin{equation}
N(y) = N_0 \cosh^{-2}(y/\lambda) + N_{\rm b},
\label{eq:Harris_den}
\end{equation}
respectively, where $B_G$ is the uniform guide magnetic field parallel to the $z$ axis, and $N_{\rm b}$ is the background uniform plasma density to demonstrate continuous plasma injection from the outside plasma sheet to the reconnection downstream.  The background plasma density $N_{\rm b}$ is set as $5 \%$ of the maximum plasma density $N(0)$ at $y=0$. The relationship between the pressure balance $B_0^2/ 8 \pi = 2 N_0 T_0$ and force balance $2 T_0/\lambda = |e| u_d B_0/c$ is satisfied, where $u_d$ is the drift velocity, and the initial electric field ${\bf E}$ is zero.

In this paper, we fix the plasma temperature $T_0/mc^2=1$, and investigate the energy partitioning between the thermal and nonthermal populations by changing only the magnitude of the guide magnetic field $B_G/B_0$ from 0 to 5/2.  The magnetization parameter $\sigma = B_0^2/(8 \pi N_0 mc^2)= 2$ if we use the number density for the value of the Harris plasma sheet $N_0$ and the magnetic field for the value of the anti-parallel magnetic field component $B_0$.
The Alfv\'{e}n speed $v_A$ is expressed as $v_A =c\sqrt{\sigma/(1+\sigma)}$
in the cold plasma limit, where the density and magnetic field are used for the values of the central Harris plasma sheet and outside plasma sheets, respectively. The inertia length $c/\omega_p$ in the plasma sheet is $6.5$ grid cell, and the ratio of the gyro-radius $r_g$ for the Harris magnetic field $B_0$ and the thickness of the plasma sheet $\lambda$ is set $r_g/\lambda=0.45$. The gyro-radius $r_g$ decreases with an increase in the guide magnetic field. The drift speed $u_d$ is set $u_d/c=0.52$.

Figures \ref{fig:structure} (a) and (b) in the top two panels show the density structures obtained for the 3D system with and without the guide magnetic field $B_G$.  The left panel (a) is the case with $B_G/B_0=1$, while the right panel (b) is the case without the guide magnetic field.  The red and blue isosurfaces are for the high and low plasma densities in the plasma sheet, respectively, and the green lines are the magnetic field lines.  Three colored contour plots outside the box show the cross section of the plasma density at $x=0$, $z/\lambda=80.6$ and $y=0$, respectively. One can clearly see that the plasma density structure becomes turbulent for the guide-field reconnection in panel (a), but not for the anti-parallel magnetic field reconnection in panel (b).  The $xy$ plane cross section above the box in panel (b) shows a large magnetic island around $x/\lambda =150$ and three small magnetic islands around $x/\lambda =20$, $220$, and $310$, while the $xy$ plane cross section in panel (a) shows a large magnetic island around $x/\lambda=200$ surrounded by many other turbulent density structures.  A similar structure can also be seen in the $yz$ plane cross section. The dynamically turbulent structure seen in the 3D guide-field reconnection is basically the same as that discussed by \citet{Daughton11}.

The bottom two panels (c) and (d) show the comparison of the plasma density between 3D and 2D guide-field reconnection with $B_G/B_0=1$.  Panel (c) is obtained by integrating the plasma density along the z-axis in panel (a), and panel (d) is the result under the same plasma parameters as panel (a) by using 2D PIC code. Comparing the density structure between the cross section (a) and the integrated structure (c), it can be seen that the patchy and turbulent density structure seen in panel (a) is more or less smeared out in the integrated density in panel (c), and the density in panel (c) is similar to the result of the 2D simulation in panel (d).  However, we will show later that the patchy density structure plays an important role in particle acceleration.

The guide magnetic field plays an important role in the patch and turbulent magnetic reconnection, as shown by comparing panels (a) and (b) obtained in 3D simulations. Comparing panels (a) and (c) for the guide-field reconnection, the 3D effect is another important condition for the patch and turbulent reconnection. The guide-field reconnection in 2D system can occur only at the neural sheet, while the guide-field reconnection in 3D system can occur not only at the neutral sheet but also outside the neutral sheet \cite{Galeev86,Daughton11,Guo21}.  This is a reason why the patch and turbulent reconnection can occur under both the effects of 3D geometry and finite guide magnetic field.

\section{Density Profile and Turbulent Waves}
\begin{figure*}
\includegraphics[width=16cm]{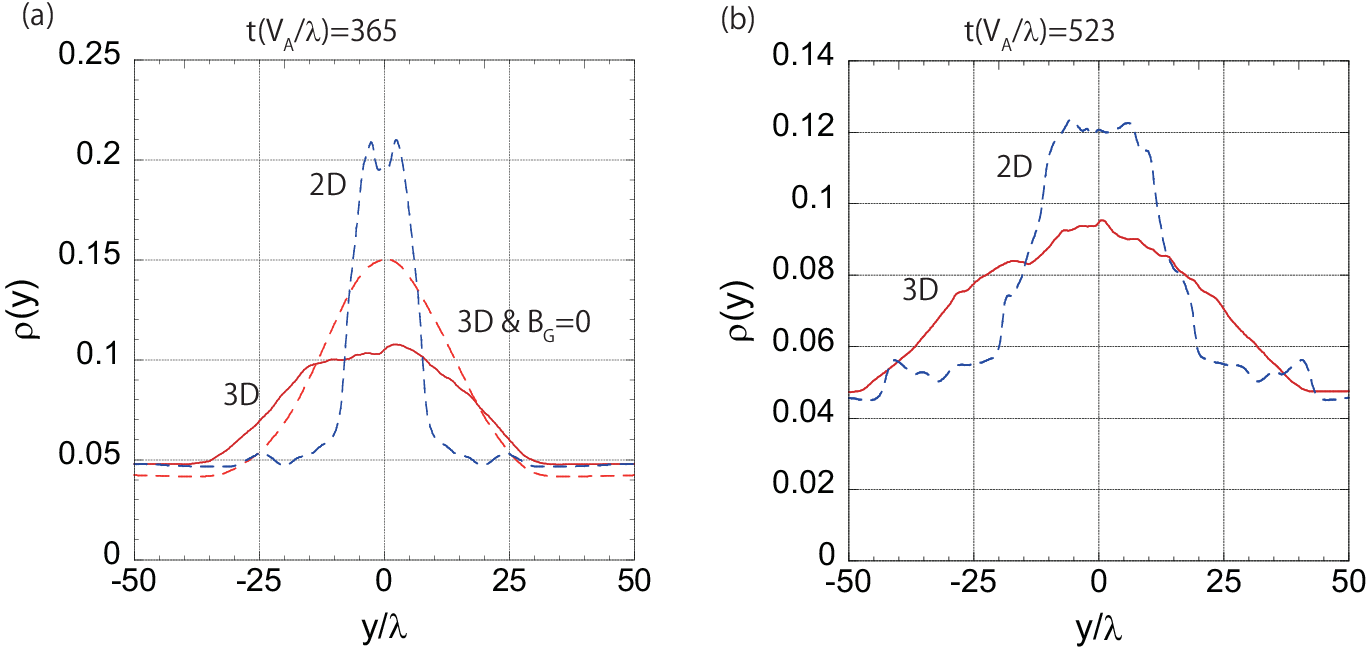}
\caption{Comparison of the density profiles of the guide-field reconnection with $B_G/B_0=1$  in 3D reconnection (red solid line) and in 2D (blue dashed line). The left panel (a) is taken at time $t/(V_A/\lambda)=365$ and the right panel (b) at $t(V_A/\lambda)=523$. For reference, the density profile of the 3D reconnection with the anti-parallel reconnection $B_G=0$ is shown as the red dashed line in the panel (a).}
\label{fig:density_profile}
\end{figure*}
It is important to \crd compare \cbk the density profile between 3D and 2D guide-field reconnection. Shown in Figures \ref{fig:density_profile} (a) and (b) are the plasma density integrated along the x-axis at $t(V_A/\lambda)=365$ and at $t(V_A/\lambda)=523$, respectively. The red solid and blue dashed curves are the 3D and 2D profiles, respectively.  With time, the high-density plasma sheet becomes thicker for both 2D and 3D reconnection, but we can clearly observe that 3D reconnection produces the wider plasma sheet compared to 2D reconnection.  The small bump/shoulder appearing around $y/\lambda \sim \pm (25-30)$ for 2D reconnection in panel (a) corresponds to the separatrix of the magnetic island.  In the later time stage at $t(V_A/\lambda)=523$, the shoulder is located around $y/\lambda \sim \pm 45$.  The positions of the foot of the high density region of the 3D reconnection are around $y/\lambda=30-35$ at $t(V_A/\lambda)=365$ and $y/\lambda=40-45$ at $t(V_A/\lambda)=523$, respectively.  The positions of these outer edges of the 3D reconnection are almost the same as the positions of the separatrix of the 2D reconnection, suggesting that the overall size of the reconnection is almost the same between 3D and 2D.

For reference, the density profile of the 3D anti-parallel magnetic field reconnection at the same time stage $t(V_A/\lambda)=365$, which is obtained from panel (b) in Figure \ref{fig:structure}, is also shown in panel (a) by the red dashed line in Figure \ref{fig:density_profile}.  Comparing the red solid line with the red dashed line, we have to keep in mind that the growth rate of the anti-parallel magnetic field reconnection is larger than that of \crd the guide magnetic field reconnection \cbk.  In fact, we find that the plasma density of the 3D anti-parallel reconnection outside plasma sheet is slightly lower than the initial plasma density $N_b/N(0)=0.05$, because the reconnection proceeds faster and the outer propagating slow mode rarefaction wave pulls the outside plasma into the central plasma sheet. Nevertheless, the density profile with the anti-parallel reconnection (red dashed line) is narrower than that with the guide magnetic field (red solid line).  In addition, the density profile around the neutral sheet shows a flat shape for the guide-field reconnection, suggesting strong diffusion in the central plasma sheet for the guide-field reconnection.
\begin{figure*}
\includegraphics[width=16cm]{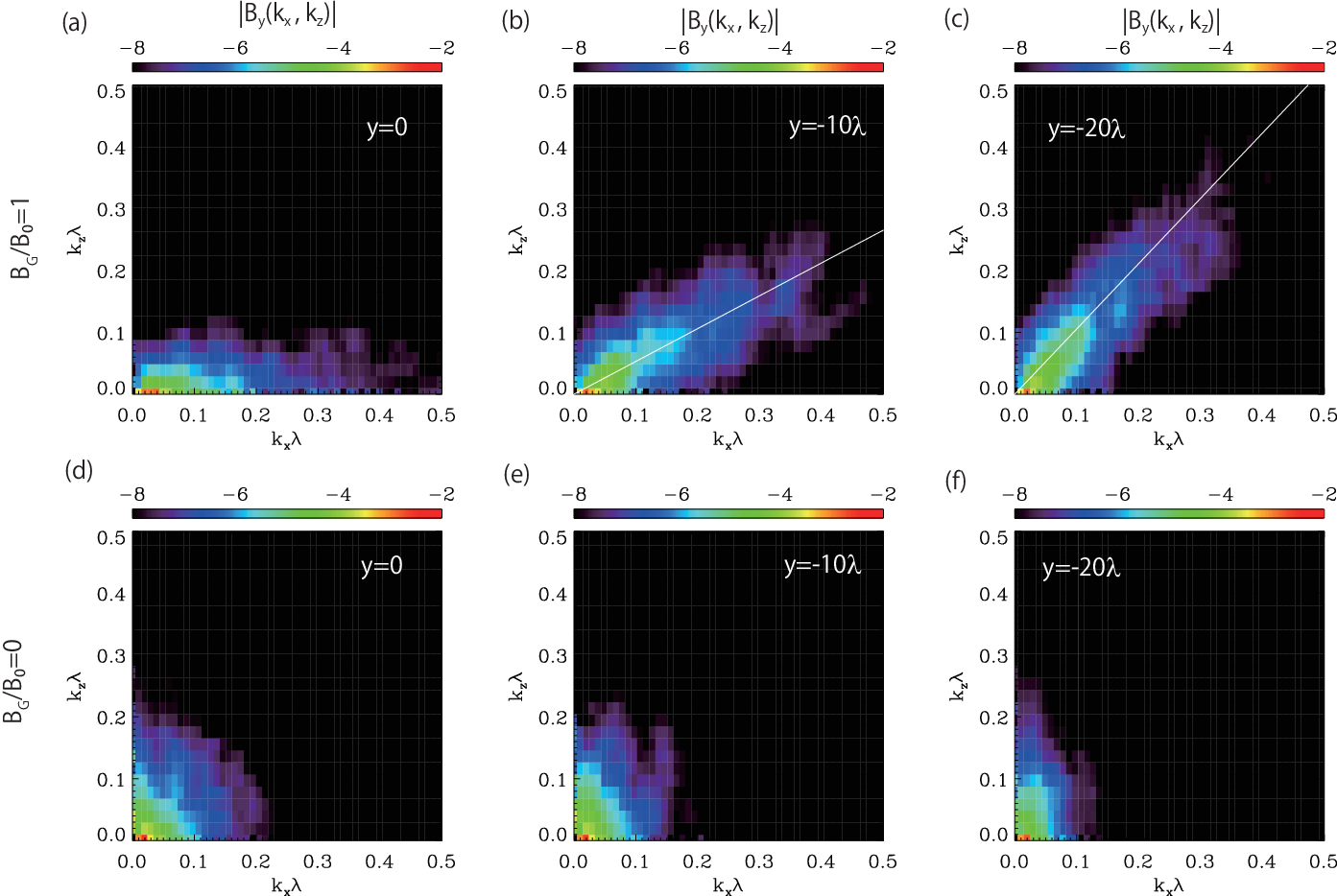}
\caption{Wave spectra $|B_y(k_x, y, k_z)|$ of oblique tearing mode with the guide magnetic field $B_G/B_0=1$, (a) at the neutral sheet $y=0$, (b) at $y/\lambda=-10$, and (c) at $y/\lambda=-20$.}
\label{fig:oblique_tearing}
\end{figure*}

The turbulent and broad density profile in the 3D guide-field reconnection may be related to the excitation of the oblique tearing mode \cite{Galeev77,Drake77,Quest81,Karimabadi05}.  It is interesting to examine the wave spectrum associated with the turbulent reconnection. While the tearing mode propagates parallel to the x-axis for the anti-parallel magnetic field reconnection, the oblique tearing mode can be excited outside the neutral sheet for the guide-field reconnection to satisfy $\bf{k} \cdot \bf{B}=0$, where $\bf{k}$ is the wave vector of the tearing mode.  Since the ambient magnetic field $\bf{B}$ is a function of $y$,  the propagation of the tearing mode also changes.  Furthermore, when the sizes of these magnetic islands due to the oblique tearing mode/reconnection become large enough to overlap, the islands are destroyed and the particles begin to wander stochastically.  This process can lead to the diffusion of the plasma density.

The top three panels (a), (b) and (c) in Figure \ref{fig:oblique_tearing} show the Fourier spectra of $B_y(k_x, y, k_z)$ obtained for the guide-field reconnection with $B_G/B_0=1$ at three different positions of $y/\lambda=0$, $-10$ and $-20$ at the time of \crd $t(V_a/\lambda)=366$ \cbk, respectively. The amplitudes are normalized to the Harris magnetic field $B_0$. The bottom three panels of (d), (e) and (f) are those for the anti-parallel magnetic field reconnection with $B_G/B_0=0$.  From the top three panels of (a), (b) and (c), we see that the excited waves with long wavelengths $k \lambda < 1$ propagate along the x-axis at the neutral sheet at $y=0$, and the waves at $y/\lambda =-10$ and $-20$ propagate obliquely away from the x-axis.  The propagation angle at $y/\lambda=-20$ is more oblique than that at $y/\lambda=-10$, and we think that these turbulent and broadband waves are generated to satisfy the resonance condition of $\bf{k} \cdot \bf{B}=0$ for the oblique tearing mode.  In fact, the white lines depicted in Figure \ref{fig:oblique_tearing} (b) and (c) are the slopes calculated from the average magnetic field magnitudes $\langle B_x(y) \rangle = \int \int B_x(x,y,z) dx dz$ and $\langle B_z(y) \rangle = \int \int B_z(x,y,z) dx dz$, indicating that the propagation angle satisfies $\bf{k} \cdot \bf{B}=0$ \cite{Daughton11}.    In contrast, the anti-parallel magnetic field reconnection in the bottom three panels of (d), (e), and (f) does not show any behavior of the oblique tearing mode. Therefore, we can understand that the patchy density structure of the guide-field reconnection in Figure \ref{fig:density_profile} (a) is due to the oblique tearing mode.

In addition to the wave spectra of $B_y(k_x, y, k_z)$ showing the oblique tearing mode excitation, we have also studied the wave spectra of the plasma density and the electric field $E_z$, which can capture other instabilities such as the drift-kink mode \cite{Zenitani05a,Zenitani07}.  The profile of these wave spectra in the nonlinear phase was basically the same as that shown in Figure \ref{fig:oblique_tearing}.  However, in the linear phase for the weak guide magnetic field cases with $B_G/B_0 < 1/4$, we clearly observed two unstable modes: one is the tearing mode with the maximum growth rate along the x-axis and the other is the drift-kink mode with the maximum growth rate along the z-axis (not shown here).

\begin{figure*}
\includegraphics[width=16cm]{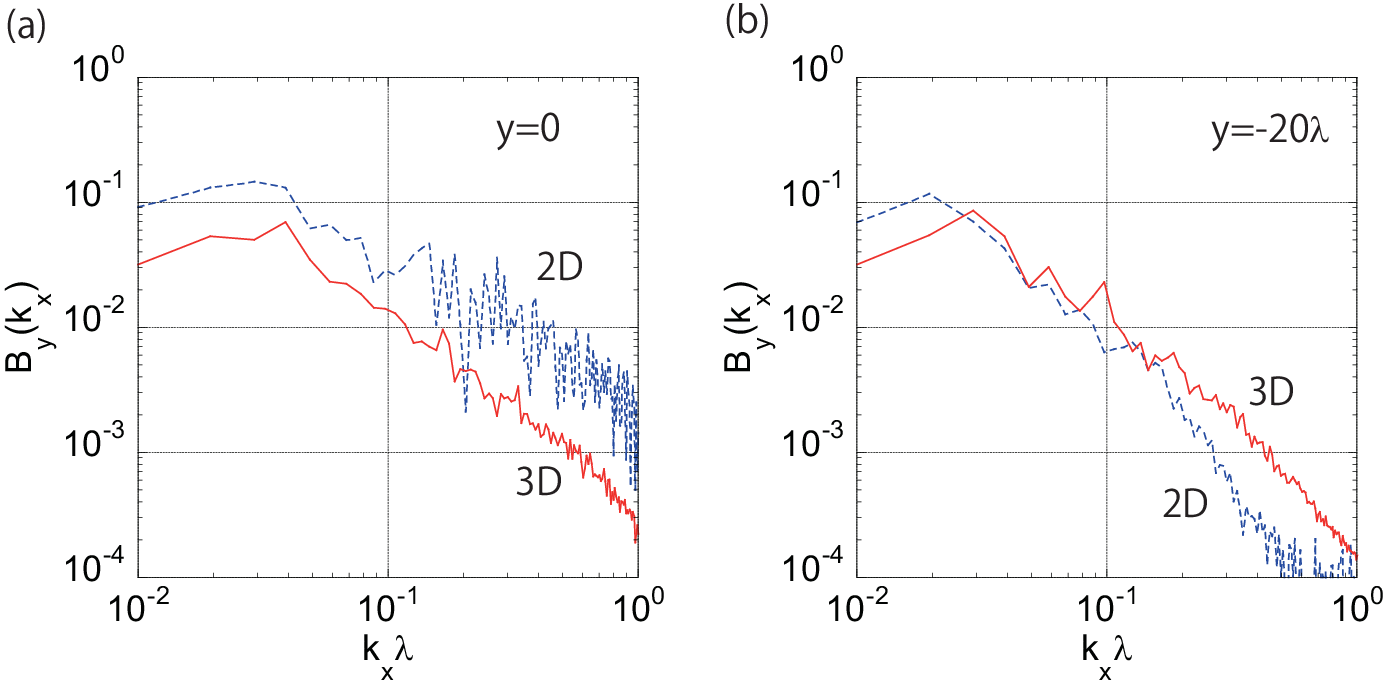}
\caption{Comparison of the wave spectra of the guide-field reconnection with $B_G/B_0=1$ in 3D reconnection (red solid line) and in 2D (blue dashed line). The left panel (a) is taken at the neutral sheet $y=0$ and the right panel (b) at $y=-20 \lambda$.}
\label{fig:turbulent_spectrum}
\end{figure*}

The intensity of these oblique waves is another key parameter controlling the formation of the broad density profile. Figure \ref{fig:turbulent_spectrum} shows the comparison of the waves spectra of $B_y(k_x)$ between the 3D guide-field reconnection in Figures \ref{fig:structure} (a) and the 2D guide-field reconnection in Figure \ref{fig:structure} (d).  The amplitudes are normalized to $B_0$. The left panel (a) and the right panel (b) are obtained at the neutral sheet $y=0$ and at the edge of the plasma sheet at $y=-20 \lambda$, respectively. The red solid lines and the blue dashed lines are the wave spectra of the 3D and 2D reconnections, respectively.  At the neutral sheet, the wave intensity of the 2D reconnection is stronger than that of the 3D reconnection. At the edge of the plasma sheet, however, the wave intensity of 2D reconnection becomes weaker, while the wave intensity of 3D reconnection at $y=-20\lambda$ remains almost at the same level of the wave intensity at $y=0$.  Therefore, we think that the broad density profile is generated by the particle scattering by the oblique tearing mode existing in a wide range of the plasma sheet for the 3D guide-field reconnection.

\section{Time Evolution of Plasma Heating}
\begin{figure*}
\includegraphics[width=16cm]{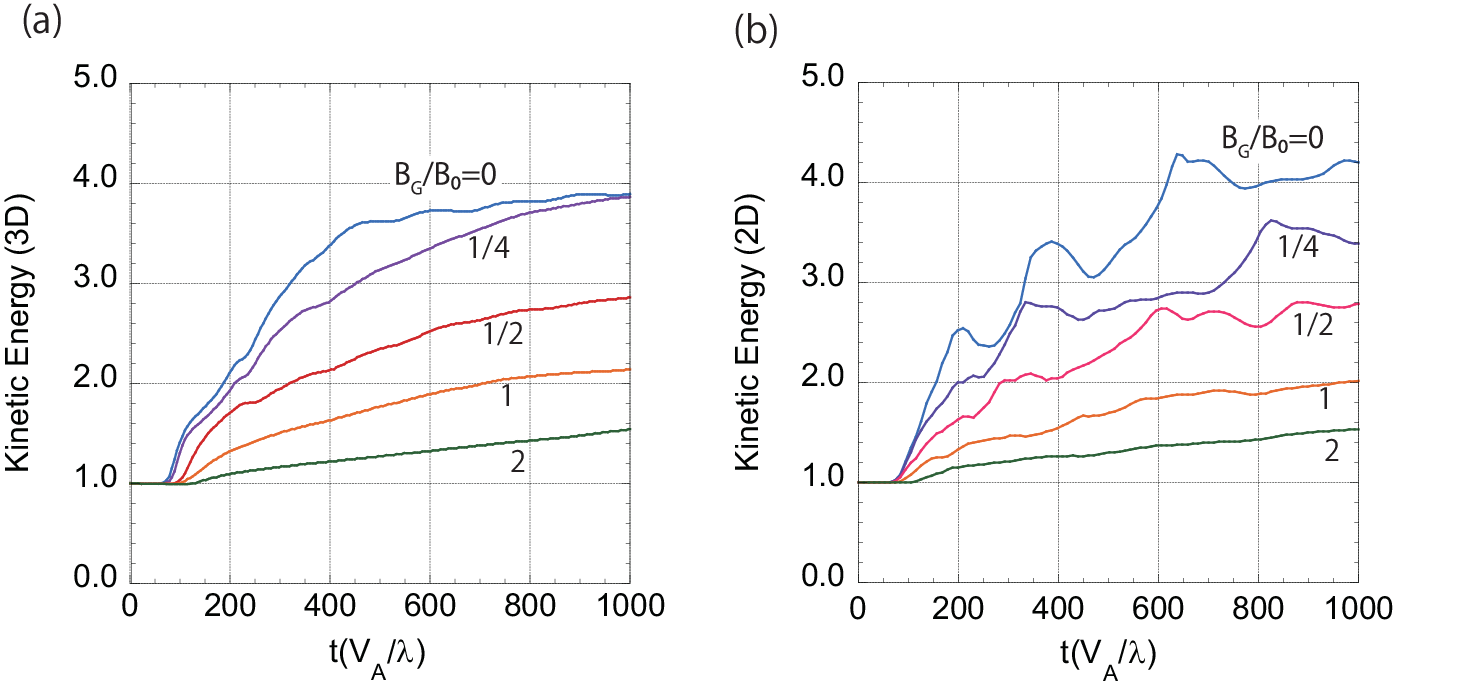}
\caption{Time evolution of the total kinetic energy in the system as a function of the guide magnetic field $B_G/B_0 = 0$, $1/4$, $1/2$, $1$ and $2$.  Panel (a) on the left and panel (b) on the right are the 3D and 2D results, respectively.}
\label{fig:energy_history}
\end{figure*}

In the previous section, we discussed that the 3D guide-field reconnection generates the dynamic and turbulent structure in the plasma sheet.  It is interesting to study whether or not the dimensionality of 3D/2D makes a difference to the plasma heating and particle acceleration.  Figures \ref{fig:energy_history} (a) and (b) are the comparison of the time history of the total kinetic energy between 3D reconnection (the left panel) and 2D reconnection (the right panel), respectively.  We have calculated the kinetic energy in the whole simulation box by summing up the individual electron and positron energies, i.e.,  $ \sum^{N_p}_{i=1}  (\gamma_i -1) m_i c^2$, where Lorentz factor $\gamma_i=1/\sqrt{1-v_i^2/c^2}$.  The total kinetic energy is normalized to the initial total kinetic energy at $t(V_A/\lambda)=0$. In our simulation study, the initial plasma parameters including the initial kinetic energy are all the same except for the magnitude of the guide magnetic field.  In Figure \ref{fig:energy_history}, five different cases for the guide-field reconnection are plotted: the solid blue, purple, red, orange, and green lines show the cases with the guide magnetic field of $B_G/B_0=0$, $1/4$, $1/2$, $1$ and $2$, respectively.

The kinetic energy at $t(V_A/\lambda)=1000$ becomes almost four times higher than that at the initial state for the case with no guide magnetic field, but the energy gain decreases with increasing the guide magnetic field for both 2D and 3D cases.  For the guide magnetic field $B_G/B_0=2$, the energy gain remains at a level of $50 \%$.  The linear growth rate of the collisionless tearing mode is known to decrease with increasing the guide magnetic field \cite{Galeev77,Drake77,Quest81,Hoshino87,Pritchett04,Karimabadi05}, because the resonant particles around the magnetic diffusion region can be easily accelerated along the finite guide magnetic field.  In fact, the resonant particles can be trapped by the guide magnetic field, and then the effective conductivity becomes high \crd \cite{Galeev77,Drake77,Quest81,Hoshino87} \cbk.  The tendency of the kinetic energy gain as a function of the guide magnetic field can be interpreted by the change in the growth rate of the oblique tearing mode.

Comparing the 3D time history in the left panel with the 2D time history in the right panel, we find several small differences: The 2D time history shows clearly several bumps associated with the coalescence of plasmoids, but the 3D time history does not show such a large oscillation in the time history of the energy.  The onset of the plasma heating occurs at almost the same time stage $t(V_A/\lambda) \sim 100$ for 2D reconnection, but the onset of 3D reconnection with the weak guide magnetic field $B_G/B_0 <1/4$ occurs earlier at $t (V_A/\lambda) \sim 70$, because the drift-kink instability can be excited in the early time stage \cite{Zenitani05a,Zenitani07,Jaroschek09}.  Although, there are some differences between 3D and 2D,  we think that the overall energy history is quite similar and there is no significant difference in the energy gain.  

\section{Model Fitting of Energy Spectrum}
\begin{figure*}
\includegraphics[width=16cm]{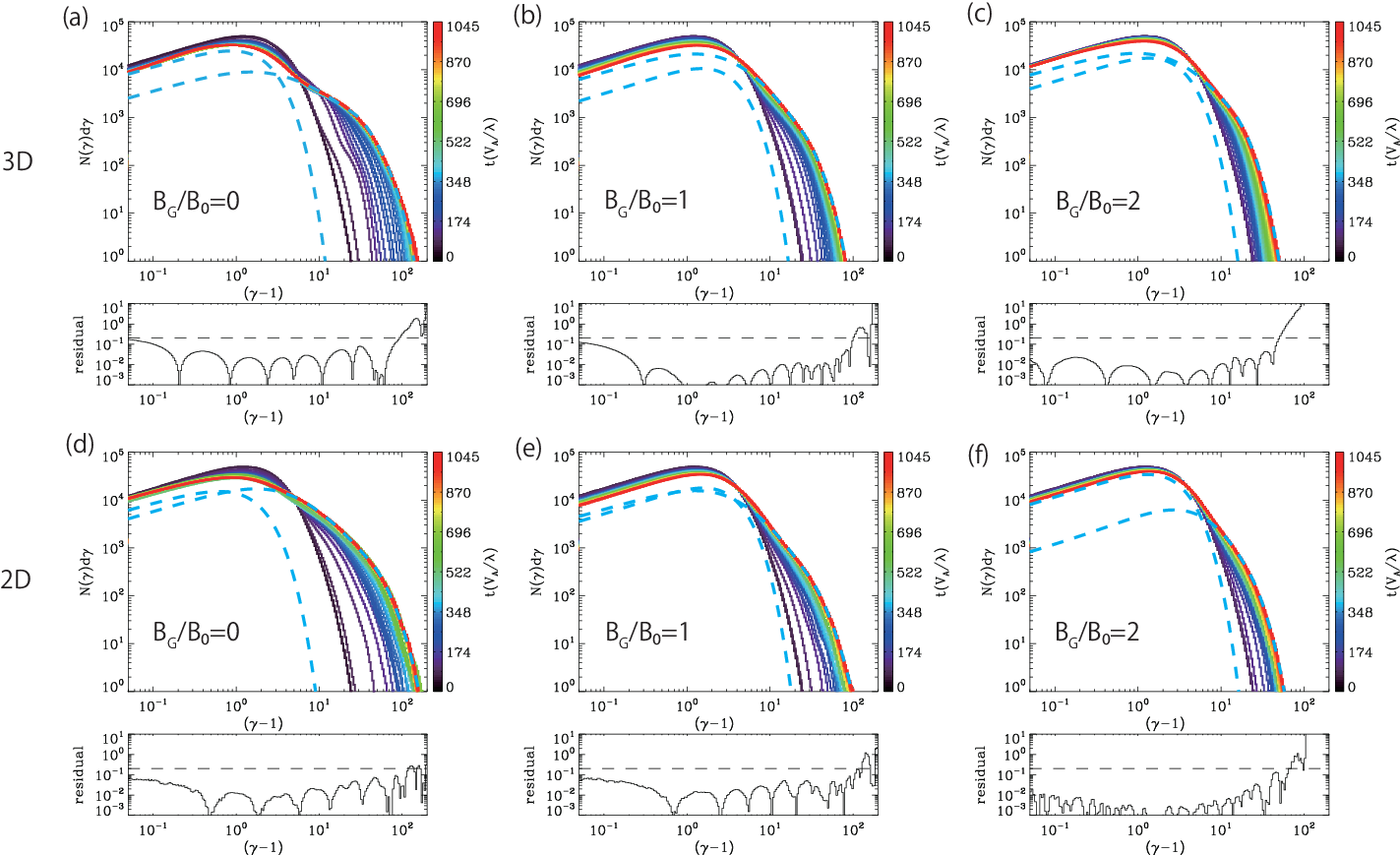}
\caption{Comparison of 2D and 3D energy spectra for the entire reconnection region. The top three panels are the 3D reconnection for (a) $B_G/B_0=0$, (b) $B_G/B_0=1$ and (c) $B_G/B_0=2$, and the bottom three panels are the 2D reconnection for (d) $B_G/B_0=0$, (e) $B_G/B_0=1$ and (f) $B_G/B_0=2$. The colored lines indicate the time evolution of the spectra, whose time stage is indicated in the right side bar.  The dark-blue line indicates the initial stage with the Maxwell distribution, and the red line indicates the final stage where a large plasmoid has formed in the system.  Two cyan dashed lines represent the model fitting result for the composite function of the Maxwell and kappa distributions.  The low-energy region corresponds to the Maxwellian part and the high energy region is the kappa distribution part.  The lower panels show the residuals of the model fitting, which indicate the difference between the model fitting and the simulation spectrum.  The dashed lines in the bottom plot may represent an acceptable residual for reference.}
\label{fig:energy_spectra}
\end{figure*}

In Paper I, we discussed the energy partitioning between thermal and nonthermal components during reconnection by using 2D simulation. Here we study it in the 3D system. Shown in Figure \ref{fig:energy_spectra} is the energy spectra for $B_G/B_0=0$, $1$ and $2$ integrated over the whole simulation domain.  In our previous paper of 2D reconnection, we studied the energy spectrum downstream of the separatrix lines sandwiched by the last reconnected magnetic field lines, but in 3D reconnection we cannot define the separatrix surface to distinguish the downstream and the upstream of the reconnected magnetic field line.  We then examine the energy spectra including both the preheated plasmas in the upstream and the heated/accelerated plasmas in the downstream.  The top three panels are the results of 3D reconnection, while the bottom three are 2D reconnection, and the colored lines illustrate the time evolution of the energy spectra.  The left two panels are the case of the anti-parallel reconnection with $B_G=0$, the middle panels are the case of the guide-field reconnection with $B_G/B_0=1$, and the right panels are the case with $B_G/B_0=2$.

The horizontal axis represents the particle kinetic energy $(\gamma -1)$ \crd normalized by $mc^2$ \cbk and the vertical axis represents the number density $N(\gamma) d\gamma$, where $\gamma=1/\sqrt{1-(v/c)^2}$. The vertical scales of $N(\gamma)d\gamma$ are shown in an arbitrary unit, but the ratio of the top to the bottom is set to five orders of magnitude.  On the horizontal axis, the number of particles is counted by logarithmic binning \crd in energy \cbk, so the one-count level is below the spectra shown here. The colored lines show the time evolution of the energy spectra, whose time stages are indicated by the colored bars on the right. The bluish and reddish colors correspond to the earlier and later stages, respectively. Since we have assumed a uniform plasma temperature in space for both the Harris plasma sheet and the background components in Equation (\ref{eq:Harris_den}), the dark blue spectra in Figure \ref{fig:energy_spectra} show the initial plasma state with the combination of the non-drifting Maxwellian distribution function and the drifting Maxwellian distribution function.  The red line is the final state when one large plasmoid is formed in the simulation box. 

Over time, we clearly observe that the Maxwellian plasmas are gradually heated by reconnection heating processes, and these heated plasmas are simultaneously further accelerated to form a nonthermal component for both 3D and 2D reconnection.  Roughly speaking, the spectra consist of the lower energy distribution close to the Maxwellian function and the higher energy population with a power-law-like distribution.  The lower energy Maxwellian-like part remains to be almost the same temperature, but the higher energy power-law-like part evolves with time.  These energy spectra for both 3D and 2D reconnections show similar time evolution.

In the late nonlinear stage at $t/(V_A/\lambda)=1045$ indicated by the red curve, we find the well-developed nonthermal energy spectra.  We now examine the spectral behavior of the red curve by using a least-squares fit to a model function. As already discussed in Paper I, the final-stage spectra can be well fitted by a composite model spectrum $N_M(\gamma) +N_{\kappa}(\gamma)$ of a Maxwell distribution function $N_M(\gamma)$ and a kappa distribution function $N_{\kappa}(\gamma)$.  The details of the model fitting function are given in Appendix.

In the six panels in Figure \ref{fig:energy_spectra}, two cyan lines are the model fitting results for \crd $N_{M+\kappa}(\gamma)=N_M(\gamma)+N_{\kappa}(\gamma)$ \cbk at the final stage.  The low energy part corresponds to the Maxwellian function $N_M(\gamma)$, and the high-energy part is the kappa distribution function $N_{\kappa}(\gamma)$. The bottom six panels below the energy spectra show the residual of the model fitting of $N_{M+\kappa}(\gamma)$ from the simulation data of $N_{data}(\gamma)$, described as
\begin{eqnarray}
{\rm residual}(\gamma) = |N_{data}(\gamma) - N_{M+\kappa}(\gamma) |/
{\rm min}(N_{data}(\gamma),N_{M+\kappa}(\gamma)),
\end{eqnarray}
where ${\rm min()}$ denotes a function that returns the minimum element from $N_{data}(\gamma)$ and $N_{M+\kappa}(\gamma)$. We observed that the residuals in wide energy ranges were less than $2 \times10^{-1}$ as indicated by the dashed lines, and we think that the model fitting of $N_{M+\kappa}(\gamma)$ described a very good model.  As the fitting curve of $N_{M+\kappa}(\gamma)$ overlaps with the red curve of the final stage, the curve of $N_{M+\kappa}(\gamma)$ is not shown here.  We confirmed that not only the above six energy spectra shown here but also other simulation results can be well approximated by the model function consisting of the Maxwellian and kappa distributions $N_{M+\kappa}(\gamma)$.  
\section{Plasma Temperature, $\kappa$ Index and Nonthermal Energy Density}
\begin{figure*}
\includegraphics[width=16cm]{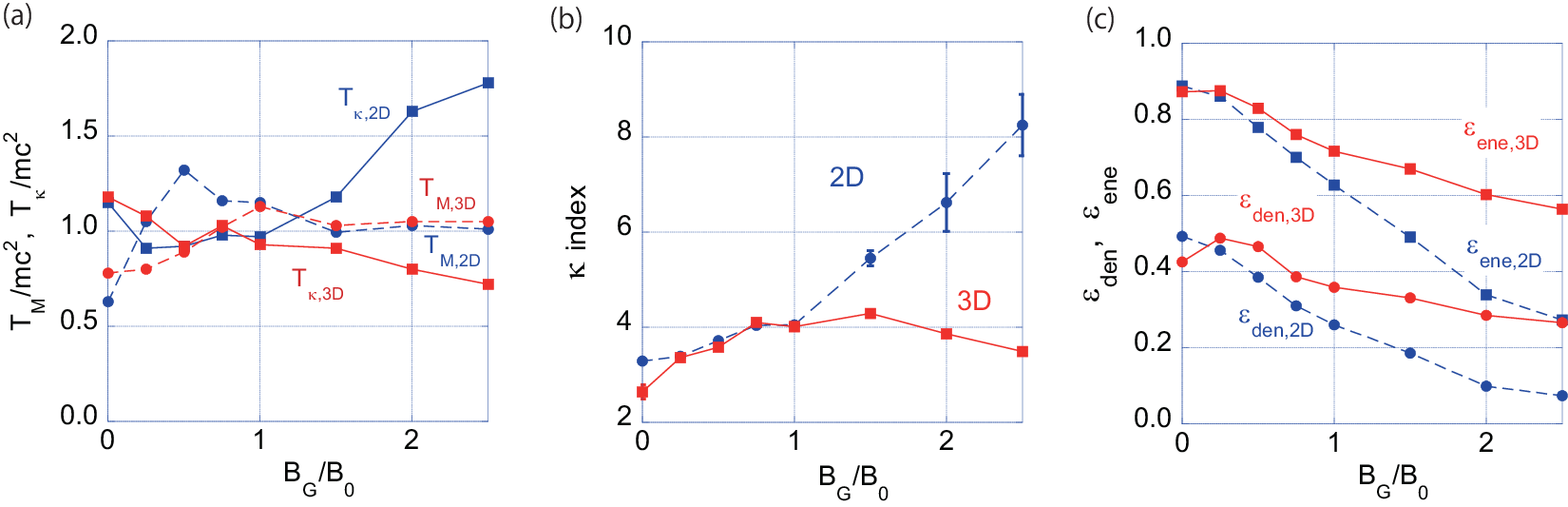}
\caption{Comparison of 3D and 2D energy spectra. (a) Maxwellian temperature $T_M$ (dashed lines) and kappa temperature $T_{\kappa}$ (solid lines), (b) kappa index for 2D (blue dashed line) and 3D (red solid line), and (c) the nonthermal energy $\varepsilon_{\rm ene}$ and the number density $\varepsilon_{\rm den}$ as a function of the guide magnetic field $B_G/B_0$.}
\label{fig:model_fitting}
\end{figure*}

Based on our model fitting results discussed above, we study the plasma heating and acceleration behavior in more detail.  The left panel (a) in Figure \ref{fig:model_fitting} shows the thermal temperatures $T_M$ and $T_{\kappa}$ normalized by the rest mass energy as the function of the guide magnetic field $B_G/B_0$.  The red solid and dashed lines represent the temperature of the Maxwellian part $T_M$ and the thermal temperature of the kappa distribution function $T_{\kappa}$ for 3D reconnection, while those with the blue lines are $T_M$ and $T_{\kappa}$ for 2D reconnection.
The 2D results with $B_G/B_0<1$ are basically the same as Paper I.
These temperatures are obtained by averaging the model fitting values in the near final stage of $t(V_A/\lambda)=914-1045$ with the reddish colors in Figure \ref{fig:energy_spectra}.  The Maxwellian temperatures for both 3D and 2D cases remain  close to $T_M/mc^2=1$, maintaining the initial plasma temperature. This component may represent the plasma situated outside the plasma sheet before heating, but some plasmas may enter into the plasma sheet without any significant heating process \cite{Hoshino23}.  More important is the behavior of the kappa temperature $T_{\kappa}$ in a strong guide magnetic field regime with $B_G/B_0 >3/2$.  While the 2D kappa temperature $T_{\kappa,2D}$ increases with increasing the guide magnetic field, the 3D kappa temperature $T_{\kappa.3D}$ remains almost the same or decreases slightly with increasing the guide magnetic field.

In addition to the thermal plasma property mentioned above, the behavior of the kappa index shown in the middle panel (b) can provide insight into the energy partitioning between the thermal and the nonthermal plasmas. The 3D and 2D results are displayed as the red solid and the blue dashed lines, respectively, as a function of the guide magnetic field.  
The error bars, which are mostly smaller than the closed squares/circles, represent one-sigma standard deviation. We find that the kappa index for 3D remains around 3-4 regardless of the guide magnetic field.  Similarly, for the 2D system, the kappa index also shows a hard spectrum for $B_G/B_0 < 1$, but increases with an increase in the guide magnetic field.  
Since the thermal temperature of the kappa distribution $T_{\kappa,2D}$ for 2D in panel (a) increases with an increasing guide magnetic field, we believe that the 2D reconnection can effectively heat the plasma rather than accelerate the nonthermal particle.

\begin{figure*}
\includegraphics[width=16cm]{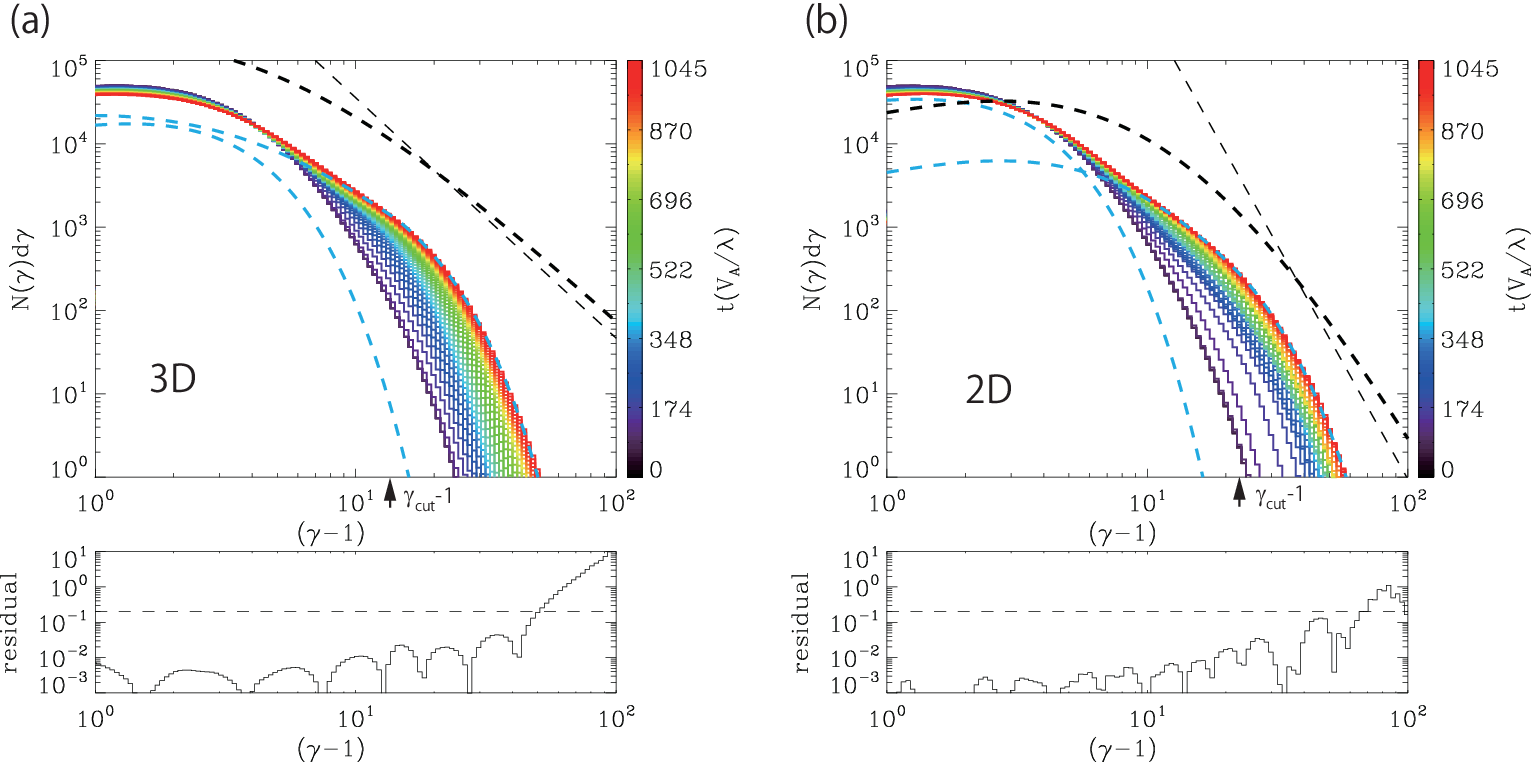}
\caption{Expanded energy spectra for panels (c) and (f) in Figure \ref{fig:energy_spectra}, and the format is basically the same as shown in Figure \ref{fig:energy_spectra}. Panels (a) and (b) are 3D and 2D results for $B_G/B_0=2$, respectively.  For reference, the thick and thin dashed line show the kappa distribution function and the power-law function, respectively.}
\label{fig:expand_spectra}
\end{figure*}

To improve the clarity of the spectral behavior in 3D and 2D, Figure \ref{fig:expand_spectra} has been expanded to show the transition energy from the thermal to the nonthermal regime in Figure \ref{fig:energy_spectra}. The left panel (a) and the right panel (b) in Figure \ref{fig:expand_spectra} correspond to panels (c) and (f) in Figure \ref{fig:energy_spectra}.  The red curves of the final state in 3D and 2D look similar, but the spectral behavior of the model fitting above was quite different. To see the difference, the thick black dashed line shows the kappa distribution function neglecting the cutoff contribution, i.e., $\gamma \sqrt{\gamma^2-1} \left(1+(\gamma-1)/(\kappa T_{\kappa}/mc^2)\right)^{-(\kappa+1)}$, and the thin dashed line represents a power-law function $\gamma^{-(\kappa-1)}$.  
\crd
The thick dashed lines in panels (a) and (b) correspond to those of the high-energy cyan dashed lines below the cutoff energies, which were $\gamma_{cut}=14.1$ in 3D and $22.9$ in 2D. In the model-fitting of the 3D result in panel (a), the number density of the kappa distribution is larger than that of the Maxwellian, and the kappa temperature was $T_{\kappa}/mc^2=0.79$, which is well separated from the the cutoff energy $\gamma_{cut}$.  Then the slope around $8 \lesssim (\gamma-1) \lesssim 30$ represents the nonthermal particle contribution, and the difference between the thick and thin curves, i.e., the kappa and the power-law distributions is minimal.
However, the slope in panel (b) of the 2D result, the slope below $(\gamma-1) \sim 15$ is still a part of the thermal contribution of the kappa distribution, because of the kappa temperature $T_{\kappa}/mc^2=1.65$.  Furthermore, since the number density of the kappa distribution is smaller than that of the Maxwellian distribution, we find that the slope indicated by the red line around $5 \lesssim (\gamma-1) \lesssim 30$ comes from both the contributions of the Maxwellian distribution of the lower-energy cyan curve and the kappa distribution of the higher-energy cyan curve.  In addition, the cutoff energy $\gamma_{cut}=22.9$ is not necessarily well separated from the kappa temperature of $T_{\kappa}/mc^2=1.65$.  These spectra illustrate the challenge of determining the nonthermal power-law index, when the separation between the thermal and the nonthermal populations is poor.
\cbk

Panel (c) in Figure \ref{fig:model_fitting} shows the fraction of the nonthermal number density ($\varepsilon_{\rm den})$ and the nonthermal energy density ($\varepsilon_{\rm ene}$) as the function of the guide magnetic field $B_G/B_0$, where $\varepsilon_{\rm den}$ and $\varepsilon_{\rm ene}$ are defined as,
\begin{eqnarray}
  \varepsilon_{\rm den}=
  \int_1^{\infty} (N_{\kappa}(\gamma)-N_{\kappa}^{\rm M}(\gamma)) d \gamma /
       \int_1^{\infty} N_{{\rm M}+\kappa}(\gamma) d \gamma,
\label{eq:efficiency_den}
\end{eqnarray}
and
\begin{eqnarray}
  \varepsilon_{\rm ene}=
  \int_1^{\infty} (\gamma -1) (N_{\kappa}(\gamma)-N_{\kappa}^{\rm M}(\gamma)) d \gamma /
       \int_1^{\infty} (\gamma -1) N_{{\rm M}+\kappa}(\gamma) d \gamma,
\label{eq:efficiency_ene}
\end{eqnarray}
where $N_{\kappa}^{\rm M}(\gamma)$ represents the fraction of the Maxwellian distribution function in the $\kappa$ distribution function (see Appendix for the definition of $N_{\kappa}^{\rm M}$). Therefore, the numerator of $N_{\kappa}(\gamma)-N_{\kappa}^{\rm M}(\gamma)$ corresponds only to the nonthermal power-law component.  In Figure \ref{fig:model_fitting} the red solid lines show the results of the 3D reconnection, while the blue dashed lines are the results of the 2D reconnection.

In the weak guide magnetic field regime of $B_B/B_0 < 1/4$, the nonthermal energy efficiency is significantly large with $\varepsilon_{\rm ene} \sim 0.9$, and the nonthermal number density $\varepsilon_{\rm den}$ is about 0.5.  However, the nonthermal efficiency for both 3D and 2D reconnections decreases with increasing the guide magnetic field $B_G$.  For example, in 2D reconnection the nonthermal energy density $\varepsilon_{\rm ene,2d}$ decreases from $0.9$ at $B_G/B_0=0$ to $0.27$ at $B_G/B_0=5/2$.  Although 3D reconnection shows the similar behavior of decreasing nonthermal particle production with increasing the guide magnetic field, the efficiency of the nonthermal production is much higher than that of 2D.  The nonthermal energy density of $\varepsilon_{\rm ene,3d}$ still holds about $0.6$.  Therefore, we find that the strong guide-field reconnection suppresses the conversion of the magnetic energy into the kinetic energy, but the nonthermal production rate against the total plasma heating is not necessarily low in 3D reconnection.

Looking at the time history of the total kinetic energy in Figure \ref{fig:energy_history}, the energies  increase gradually with time.  At the final time stage at $t(V_A/\lambda)=1045$, however, we have confirmed that one large magnetic island is formed in the system even for the strong guide magnetic field of $B_G/B_0=2$ and $5/2$.  In addition, from the time history of the energy spectra in Figure \ref{fig:energy_spectra}, we think that the final spectra shown by the red curves seem to reach the almost final stable state. Therefore, the nonthermal efficiencies of $\varepsilon_{\rm den}$ and $\varepsilon_{\rm ene}$ estimated from the energy spectra are almost the final states in our simulation box.  We conclude that 3D reconnection can produce more nonthermal particles compared to 2D reconnection and can maintain a hard energy spectrum even for a strong guide-field reconnection.

\begin{figure*}
\includegraphics[width=16cm]{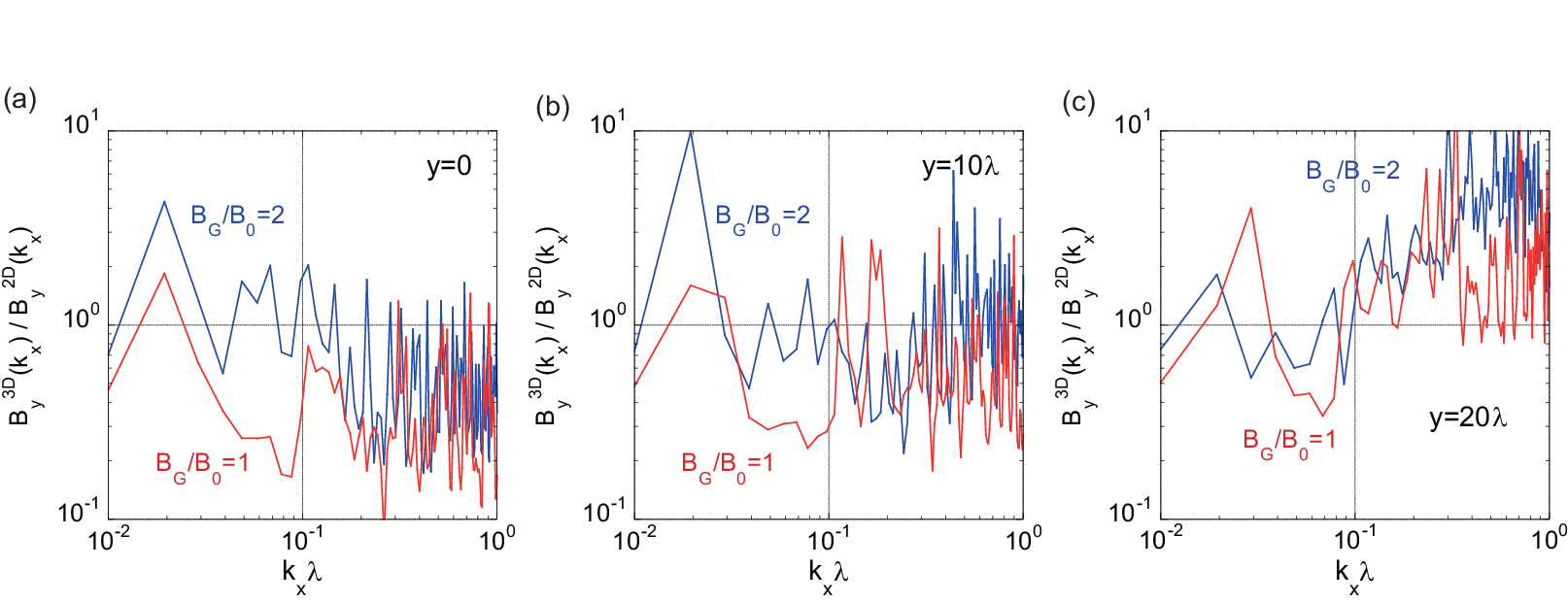}
\caption{\crd Comparison of the ratio of the 3D to 2D wave spectra as a function of the wavenumber $k_x \lambda$. The red solid line is the case of the guide-field $B_g/B_0=1$ and the blue solid line is the case of $B_g/B_0=2$.  The left panel (a) is taken at the neutral sheet $y=0$, the middle panel (b) at $y=10 \lambda$, and the right panel (c) at $y=20 \lambda$. \cbk}
\label{fig:spectra_ratio}
\end{figure*}
\crd We discussed that the patch and turbulent waves generated in 3D reconnections can enhance the efficiency of the nonthermal particle acceleration and that a hard nonthermal energy spectrum can be maintained even in the presence of a strong magnetic field.  In order to study the effect of the plasma turbulence for a strong guide-field reconnection, we show the comparison of the ratio of the wave spectra of 3D to 2D reconnection for two cases of the guide-filed $B_G/B_0=1$ and $2$ in Figure 9.  The vertical axis in Panels (a), (b), and (c) shows $B^{3D}_y(k_x)/B^{2D}_y(k_x)$ for the cases of $B_G/B_0=1$ (red line) and $B_G/B_0=2$ (blue line).  Panels (a), (b), and (c) show the ratio at the neutral sheet $y=0$, at $y=10 \lambda$, and at $y=20 \lambda$, respectively.  The ratios of $B_b/B_0=1$ and $B_G/B_0=2$ are taken at $t(V_A/\lambda)=522$ and at $t(V_A/\lambda)=914$, when the total electric field energy integrated in the whole system becomes almost maximum. The peak electric field energies normalized by the initial kinetic energy were $0.1-0.25$. For the case of $B_G/B_0=1$ at the neutral sheet $y=0$, the amplitudes of 3D spectrum is smaller than those of 2D spectrum, because the ratio is smaller than unity. At the edge of the plasma sheet $y=20 \lambda$ in Panel (c), the 3D amplitudes becomes larger than 2D amplitudes in a wide range of the wavenumber. The behavior is consistent with Panels (a) and (b) in Figure 4, even though the time stages of the spectra are not necessarily same. Let us look at the case of $B_G/B_0=2$ (blue line).  We find that the 3D turbulence of the $B_G/B_0=2$ is stronger than that of $B_G/B_0=1$ from the neutral sheet at $y=0$ to the edge of the reconnecting plasma sheet at $y=20 \lambda$.  This result supports the importance of the turbulence in the 3D guide-field reconnection. \cbk
\section{Discussions and Conclusion}
In the present work, we have investigated the efficiency of the nonthermal particle acceleration during a relativistic reconnection with a guide magnetic field by using 3D particle-in-cell simulations, and have compared the efficiency with that obtained from 2D simulations.  For simplicity, we assumed a pair plasma with electrons and positrons, and the initial plasma temperature is fixed to be $T_0/mc^2=1$.  Under this assumption, we studied how the nonthermal particle acceleration is modified as the function of the guide magnetic field $B_G$.  So far the particle acceleration process during reconnection has been extensively investigated by many authors, but these studies are limited to the cases with the anti-parallel magnetic field or with a weak guide magnetic field of \crd $B_G/B_0 \lesssim 1$ \cite{Daughton11,Dahlin17,Li19,Guo21,Zhang21} \cbk. In this paper, we have studied the strong guide-field reconnection up to $B_G/B_0 =5/2$.  Our simulation study confirmed that the efficiency of the nonthermal particle acceleration decreases with the increase of the guide magnetic field as expected before, but more importantly the energy spectrum with a strong guide magnetic field in the 3D simulation shows a hard energy spectrum with the kappa index of $\kappa \sim 4$.  This is contracted to the 2D system where the spectral index increases with increasing the guide magnetic field, and the kappa index becomes about $8$ for $B_G/B_0=5/2$.

\crd
In this paper, we have studied the case of $T_0/mc^2=1$ and $\sigma=2$.  It is important to study the $\sigma$ dependence from the nonrelativistic to the relativistic regime, but in our previous studies of two-dimensional reconnection \cite{Hoshino22,Hoshino23}, the efficiency of the plasma heating and the particle acceleration does not change in the range of $1<\sigma<10$. 
\cbk

The 3D guide-field reconnection inherently has the nature of the patch and turbulent reconnection, as multiple reconnections can occur from the neutral sheet to the outer plasma sheet.  Over time, the magnetic islands generated in the different locations can overlap and the coherent island structure seen in the 2D system can be destroyed \citep{Daughton11}.  The turbulent wave spectrum can form over a wide region of the plasma sheet.  These turbulent wave can enhance the efficiency of the nonthermal acceleration by the 2nd-order Fermi acceleration process.  If the coherent structure of the magnetic island still remains after the overlapping of the multiple islands, the multiple scattering among the magnetic islands can generate the nonthermal particle through the 1st-order Fermi acceleration process \cite{Hoshino12a}. In any case, the generation of the patchy and turbulent reconnection can improve the efficiency of the particle acceleration.

We have discussed the efficiency of particle acceleration enhanced by the patchy and turbulent nature of 3D reconnection as one of the additional acceleration processes, and \crd we have not necessarily discussed the details of the particle acceleration mechanisms that have been studied by many people before. The main acceleration in relativistic reconnection discussed earlier \cbk is the direct acceleration in and around the magnetic diffusion region, because the relative size of the magnetic diffusion region with $E>B$ supported by an inductive electric field is not necessarily small for a relativistic reconnection \cite{Zenitani01}.  In addition to this direct acceleration, the particles ejected from the diffusion region collide with the magnetic field pile-up region formed in front of the magnetic island, and the particles may be accelerated by the gradient B drift and/or the magnetic curvature drift effect \cite{Hoshino01}.  During the nonlinear time evolution, the particles trapped inside the magnetic islands can be accelerated by the 1st-order Fermi process in conversing reconnection flows \cite{Drake06}.  During the magnetic coalescence, a secondary diffusion region appears at the interface of two magnetic islands with a stronger reconnection electric field, and then the direct acceleration can occur \cite{Oka10}.  Other various types of plasma instabilities developed in the plasma sheet can contribute to the energization of particles.  Recently, the disruption of the magnetic flux tube by the kink instability in the 3D reconnection with a weak guide magnetic field is pointed out as another acceleration process \cite{Guo21}.

We have discussed that 3D reconnection maintains a hard spectrum with $\kappa \sim 4$, even though the kappa index of 2D guide-field reconnection increases with increasing the guide magnetic field.  As we can see from the energy spectra in Figure \ref{fig:expand_spectra}, however, the separation of the nonthermal cutoff energy of $f_{cut}$ from the thermal energy is not large enough for the strong guide magnetic field $B_G/B_0=2$.  We need \crd to carefully analyse \cbk the spectral index by paying attention to the separation between the thermal and the nonthermal parts. In our model fitting, we have distinguished the nonthermal kappa distribution function from the thermal Maxwellian function, and we believe that our analysis has been done carefully under the assumption of the model fitting function of $N_{M+\kappa}$. We think that it would be better to verify this result by using a more larger simulation system where we can capture the evolution of the nonthermal component. To improve the dynamic energy range, we will probably need an extremely large supercomputer memory with a \crd system \cbk size $10^3$ times larger in a 3D simulation.

\begin{acknowledgments}
This work was supported by JSPS Grant-in-Aid for Scientific Research (KAKENHI) (Grant no. 20K20908). The author would like to thank T. Amano, S. Totorica and S. Zenitani for their valuable discussions. This work used the computational resources of  the Fugaku supercomputer provided by the RIKEN Center for Computational Science (Project ID: hp230073).
\end{acknowledgments}
\section*{DATA AVAILABILITY}
The data support the findings of this study are available from the corresponding author upon reasonable request.

\appendix
\section{Model Fitting}
In our model fitting, the one-dimensional Levenberg-Marquardt least-squares fitting program ``MPFIT'' available in IDL was used \cite{Markwardt09}, after transforming a three-dimensional distribution function into a one-dimensional distribution in the simulation frame as a function of the particle energy $\gamma$.  A logarithmic binning in the $\gamma$ axis is used to reduce the statistical error for the high energy regime.

As already discussed in Paper I, the final-stage spectra can be fitted by a composite model spectrum $N_{M+\kappa}(\gamma)$ of a Maxwell distribution function $N_M(\gamma)$ and a kappa distribution function $N_{\kappa}(\gamma)$ \citep{Vasyliunas68}, as described below.
\begin{equation}
N_{M+\kappa}(\gamma) = N_M(\gamma) +N_{\kappa}(\gamma),
\label{eq:modelfunction}
\end{equation}
where
\begin{equation}
N_M(\gamma) =n_M \gamma \sqrt{\gamma^2-1} \exp(-\frac{\gamma-1}{T_M/mc^2}),
\end{equation}
and
\begin{equation}
N_{\kappa}(\gamma) =n_{\kappa} \gamma \sqrt{\gamma^2-1}
\left(1 + \frac{\gamma-1}{\kappa T_{\kappa}/mc^2} \right)^{-(1+\kappa)} f_{cut}(\gamma),
\label{eq:kappa_distribution}
\end{equation}
where $f_{cut}(\gamma)$ represents a possible high-energy cutoff in the simulation data \citep{Oka15,Werner16,Petropoulou18}, which is described as
\begin{equation}
f_{cut}(\gamma) = \left\{
\begin{array}{@{\,}ll}
1 & \mbox{for $\gamma \le \gamma_{cut}$ }, \\
\exp\left(-(\gamma -\gamma_{cut})^s/\gamma_{cut}^s \right) & \mbox{for $\gamma > \gamma_{cut}$ }.
\end{array}
\right.
\end{equation}
A high-energy cutoff can arise from the finite-time evolution of reconnection under a limited system size \citep{Petropoulou18}. Although the high-energy cutoff is important for understanding the overall energy budget of the nonthermal component, the dominant contribution of the nonthermal energy density comes from the low-energy component as long as $\kappa > 3$.  The power index $s$ of the cutoff function $f_{cut}$ is a free parameter, but it does not affect the estimation of $\varepsilon_{den}$ and $\varepsilon_{ene}$.

To distinguish thermal plasma heating from nonthermal particle acceleration in our model fitting, we introduce the idea of the Maxwellian part of the kappa distribution function $N_{\kappa}^{\rm M}(\gamma)=\lim_{\kappa \to \infty} N_{\kappa}(\gamma)$. The kappa distribution consists of a thermal Maxwellian distribution at low energies and a nonthermal population approximated by a power-law function at high energies with $N_{\kappa}(\gamma) \propto \gamma^{-\kappa+1}$, where a power-law index $s=\kappa-1$ for $\gamma\gg T_{\kappa}/mc^2$. 
We defined the thermal part of the kappa distribution by taking the limit of $\kappa \to \infty$, keeping the other fitted parameters the same, because a kappa distribution function approaches a Maxwellian distribution as $\kappa \to \infty$, as described below:
\begin{eqnarray}
\lim_{\kappa \to \infty} \left(1 + \frac{\gamma-1}{\kappa T_{\kappa}/mc^2} \right)^{-(\kappa+1)} \simeq
    \exp \left( - \frac{\gamma -1}{T_{\kappa}/mc^2} \right).
\end{eqnarray}



\begin{thebibliography}{}
\bibitem[Blackman \& Field(1994)]{Blackman94} Blackman, E.~G. \& Field, G.~B.\ 1994, \prl, 72, 494. doi:10.1103/PhysRevLett.72.494
\bibitem[Birn \& Priest(2007)]{Birn07} Birn, J. \& Priest, E.~R.\ 2007, Reconnection of magnetic fields : magnetohydrodynamics and collisionless theory and observations / edited by J. Birn and E. R. Priest. Cambridge : Cambridge University Press, 2007. ISBN: 9780521854207 (hbk.)
\bibitem[Zweibel \& Yamada(2009)]{Zweibel09} Zweibel, E.~G. \& Yamada, M.\ 2009, \araa, 47, 291. doi:10.1146/annurev-astro-082708-101726
\bibitem[Hoshino \& Lyubarsky(2012)]{Hoshino12b} Hoshino, M., \& Lyubarsky, Y.\ 2012, Space Science Reviews, 173, 521
\bibitem[Uzdensky(2016)]{Uzdensky16} Uzdensky, D.~A.\ 2016, Magnetic Reconnection: Concepts and Applications, 473
  
\bibitem[Lin et al.(2003)]{Lin03} {Lin}, R.~P., {Krucker}, S., {Hurford}, G.~J., {Smith}, D.~M., {Hudson}, H.~S., {Holman}, G.~D., {Schwartz}, R.~A., {Dennis}, B.~R., {Share}, G.~H., {Murphy}, et al.\ 2003, \apjl, 595, L69. doi:10.1086/378932

\bibitem[Oka et al.(2018)]{Oka18} {Oka}, M., {Birn}, J., {Battaglia}, M., {Chaston}, C.~C., {Hatch}, S.~M., {Livadiotis}, G., {Imada}, S., {Miyoshi}, Y., {Kuhar}, M., {Effenberger}, et al.\ 2018, \ssr, 214, 82. doi:10.1007/s11214-018-0515-4
  
\bibitem[Baker \& Stone(1977)]{Baker77} Baker, D.~N. \& Stone, E.~C.\ 1977, \jgr, 82, 1532. doi:10.1029/JA082i010p01532
\bibitem[Hoshino et al.(2001)]{Hoshino01} Hoshino, M., Mukai, T., Terasawa, T., \& Shinohara, I.\ 2001, \jgr, 106, 25979. doi:10.1029/2001JA900052
\bibitem[{\O}ieroset et al.(2002)]{Oieroset02}  {{\O}ieroset}, M., {Lin}, R.~P., {Phan}, T.~D., {Larson}, D.~E. \& {Bale}, S.~D. \ 2002, \prl, 89, 195001. doi:10.1103/PhysRevLett.89.195001

\bibitem[Lyutikov(2003)]{Lyutikov03} Lyutikov, M.\ 2003, \mnras, 346, 540. doi:10.1046/j.1365-2966.2003.07110.x
\bibitem[Kirk(2004)]{Kirk04} Kirk, J.~G.\ 2004, \prl, 92, 181101. doi:10.1103/PhysRevLett.92.181101
\bibitem[Madejski \& Sikora(2016)]{Madejski16} Madejski, G. (Greg) . \& Sikora, M.\ 2016, \araa, 54, 725. doi:10.1146/annurev-astro-081913-040044
\bibitem[Blandford et al.(2017)]{Blandford17} Blandford, R., Yuan, Y., Hoshino, M., \& Sironi, L.,\ 2017, Space Science Reviews, 207, 291

\bibitem[Zenitani \& Hoshino(2001)]{Zenitani01} Zenitani, S., \& Hoshino, M.\ 2001, \apjl, 562, L63
\bibitem[Zenitani \& Hoshino(2005b)]{Zenitani05b} Zenitani, S. \& Hoshino, M.\ 2005, \prl, 95, 095001. doi:10.1103/PhysRevLett.95.095001
\bibitem[Zenitani \& Hoshino(2008)]{Zenitani08} Zenitani, S. \& Hoshino, M.\ 2008, \apj, 677, 530. doi:10.1086/528708
\bibitem[Jaroschek \& Hoshino(2009)]{Jaroschek09} Jaroschek, C.~H. \& Hoshino, M.\ 2009, \prl, 103, 075002. doi:10.1103/PhysRevLett.103.075002\bibitem[Sironi \& Spitkovsky(2011)]{Sironi11} Sironi, L. \& Spitkovsky, A.\ 2011, \apj, 726, 75. doi:10.1088/0004-637X/726/2/75
\bibitem[Liu et al.(2011)]{Liu11} {Liu}, Wei, {Li}, Hui, {Yin}, Lin, {Albright}, B.~J., {Bowers}, K.~J. and {Liang}, Edison P. \ 2011, Physics of Plasmas, 18, 052105. doi:10.1063/1.3589304
\bibitem[Cerutti et al.(2012a)]{Cerutti12} Cerutti, B., Uzdensky, D.~A., \& Begelman, M.~C.\ 2012, \apj, 746, 148. doi:10.1088/0004-637X/746/2/148
\bibitem[Cerutti et al.(2013)]{Cerutti13} {Cerutti}, B., {Werner}, G.~R., {Uzdensky}, D.~A. \& {Begelman}, M.~C. \ 2013, \apj, 770, 147. doi:10.1088/0004-637X/770/2/147
\bibitem[Cerutti et al.(2014)]{Cerutti14} {Cerutti}, B., {Werner}, G.~R., {Uzdensky}, D.~A. \& {Begelman}, M.~C. \ 2014, \apj, 782, 104. doi:10.1088/0004-637X/782/2/104
\bibitem[Bessho \& Bhattacharjee(2012)]{Bessho12} Bessho, N. \& Bhattacharjee, A.\ 2012, \apj, 750, 129. doi:10.1088/0004-637X/750/2/129
\bibitem[Sironi et al.(2014)]{Sironi14} Sironi, L., \& Spitkovsky, A.\ 2014, \apjl, 783, L21
\bibitem[Guo et al.(2014)]{Guo14} Guo, F., Li, H., Daughton, W., \& Liu, Y.-H.\ 2014, \prl, 113, 155005
\bibitem[Rowan et al.(2017)]{Rowan17} Rowan, M.~E., Sironi, L., \& Narayan, R.\ 2017, \apj, 850, 29. doi:10.3847/1538-4357/aa9380
\bibitem[Werner et al.(2018)]{Werner18} {Werner}, G.~R., {Uzdensky}, D.~A., {Begelman}, M.~C., {Cerutti}, B. \& {Nalewajko}, K. \ 2018, \mnras, 473, 4840. doi:10.1093/mnras/stx2530
\bibitem[Ball et al.(2018)]{Ball18} Ball D., Sironi L., {\"O}zel F., 2018, \apj, 862, 80. doi:10.3847/1538-4357/aac820
\bibitem[Totorica et al.(2020)]{Totorica20} {Totorica}, S.~R., {Hoshino}, M., {Abel}, T. \& {Fiuza}, F.
\ 2020, Physics of Plasmas, 27, 112111. doi:10.1063/5.0021169
\bibitem[French et al.(2023)]{French23} {French}, Omar, {Guo}, Fan, {Zhang}, Qile \& {Uzdensky}, Dmitri A.  \ 2023, \apj, 948, 19. doi:10.3847/1538-4357/acb7dd
  
\bibitem[Eastwood et al.(2013)]{Eastwood13} {Eastwood}, J.~P., {Phan}, T.~D., {Drake}, J.~F., {Shay}, M.~A., {Borg}, A.~L., {Lavraud}, B. \& {Taylor}, M.~G.~G.~T. \ 2013, \prl, 110, 225001. doi:10.1103/PhysRevLett.110.225001
\bibitem[Hoshino(2018)]{Hoshino18} Hoshino, M.\ 2018, \apjl, 868, L18. doi:10.3847/2041-8213/aaef3a
\bibitem[Hoshino(2022)]{Hoshino22} Hoshino, M.\ 2022, Physics of Plasmas, 29, 042902. doi:10.1063/5.0086316
\bibitem[Hoshino(2023)]{Hoshino23} Hoshino, M.\ 2023, \apj, 946, 77. doi:10.3847/1538-4357/acbfb5

\bibitem[Galeev et al.(1986)]{Galeev86} Galeev, A.~A., Kuznetsova, M.~M., \& Zelenyi, L.~M.\ 1986, \ssr, 44, 1. doi:10.1007/BF00227227
\bibitem[Daughton et al.(2011)]{Daughton11} {Daughton}, W., {Roytershteyn}, V., {Karimabadi}, H., {Yin}, L., {Albright}, B.~J., {Bergen}, B. \& {Bowers}, K.~J. \ 2011, Nature Physics, 7, 539. doi:10.1038/nphys1965
\crd
\bibitem[Dahlin et al.(2017)]{Dahlin17} Dahlin, J.~T., Drake, J.~F., \& Swisdak, M.\ 2017, Physics of Plasmas, 24, 092110. doi:10.1063/1.4986211
\cbk
\bibitem[Guo et al.(2021)]{Guo21} {Guo}, Fan, {Li}, Xiaocan, {Daughton}, William, {Li}, Hui, {Kilian}, Patrick, {Liu}, Yi-Hsin, {Zhang}, Qile \& {Zhang}, Haocheng \ 2021, \apj, 919, 111. doi:10.3847/1538-4357/ac0918
  
\bibitem[Hoshino(2015)]{Hoshino15} Hoshino, M.\ 2015, \prl, 114, 061101. doi:10.1103/PhysRevLett.114.061101
\bibitem[Harris(1962)]{Harris62} Harris, E.~G.\ 1962, Il Nuovo Cimento, 23, 115

\bibitem[Galeev \& Zeleny{\v{i}}(1977)]{Galeev77} Galeev, A.~A. \& Zeleny{\v{i}}, L.~M.\ 1977, Soviet Journal of Experimental and Theoretical Physics Letters, 25, 380
\bibitem[Drake \& Lee(1977)]{Drake77} Drake, J.~F. \& Lee, Y.~C.\ 1977, Physics of Fluids, 20, 1341. doi:10.1063/1.862017
\bibitem[Quest \& Coroniti(1981)]{Quest81} Quest, K.~B. \& Coroniti, F.~V.\ 1981, \jgr, 86, 3299. doi:10.1029/JA086iA05p03299
\bibitem[Hoshino(1987)]{Hoshino87} Hoshino, M.\ 1987, \jgr, 92, 7368. doi:10.1029/JA092iA07p07368
\bibitem[Pritchett \& Coroniti(2004)]{Pritchett04} Pritchett, P.~L. \& Coroniti, F.~V.\ 2004, Journal of Geophysical Research (Space Physics), 109, A01220. doi:10.1029/2003JA009999
\bibitem[Karimabadi et al.(2005)]{Karimabadi05} Karimabadi, H., Daughton, W., \& Quest, K.~B.\ 2005, Journal of Geophysical Research (Space Physics), 110, A03214. doi:10.1029/2004JA010749
  
\bibitem[Zenitani \& Hoshino(2005a)]{Zenitani05a} Zenitani, S. \& Hoshino, M.\ 2005, \apjl, 618, L111. doi:10.1086/427873  
\bibitem[Zenitani \& Hoshino(2007)]{Zenitani07} Zenitani, S. \& Hoshino, M.\ 2007, \apj, 670, 702. doi:10.1086/522226

\bibitem[Li et al.(2019)]{Li19} {Li}, Xiaocan, {Guo}, Fan, {Li}, Hui, {Stanier}, Adam \&
  {Kilian}, Patrick \ 2019, \apj, 884, 118. doi:10.3847/1538-4357/ab4268sRevLett.127.185101
\bibitem[Zhang et al.(2021)]{Zhang21} {Zhang}, Qile, {Guo}, Fan, {Daughton}, William, {Li}, Hui \& {Li}, Xiaocan \ 2021, \prl, 127, 185101. doi:10.1103/PhysRevLett.127.185101
\bibitem[Oka et al.(2010)]{Oka10} {Oka}, M., {Phan}, T. -D., {Krucker}, S., {Fujimoto}, M. \& {Shinohara}, I. \ 2010, \apj, 714, 915. doi:10.1088/0004-637X/714/1/915
\bibitem[Hoshino(2012)]{Hoshino12a} Hoshino, M.\ 2012, \prl, 108, 135003. doi:10.1103/PhysRevLett.108.135003

\bibitem[Pritchett(2001)]{Pritchett01} Pritchett, P.~L.\ 2001, \jgr, 106, 3783. doi:10.1029/1999JA001006
\bibitem[Drake et al.(2006)]{Drake06} {Drake}, J.~F., {Swisdak}, M., {Che}, H. \& {Shay}, M.~A. \ 2006, \nat, 443, 553. doi:10.1038/nature05116
\bibitem[Guo et al.(2020)]{Guo20} Guo, F., Liu, Y.-H., Li, X., Li, H., Daughton, W., \& Kilian, P.\ 2020, Physics of Plasmas, 27, 080501. doi:10.1063/5.0012094
  
\bibitem[Markwardt(2009)]{Markwardt09} Markwardt, C.~B.\ 2009, Astronomical Data Analysis Software and Systems XVIII, 411, 251. doi:10.48550/arXiv.0902.2850
\bibitem[Vasyliunas(1968)]{Vasyliunas68} Vasyliunas, V.~M.\ 1968, \jgr, 73, 2839. doi:10.1029/JA073i009p02839
\bibitem[Oka et al.(2015)]{Oka15} {Oka}, Mitsuo \& {Krucker}, S{\"a}m \& {Hudson}, Hugh S. \& {Saint-Hilaire}, Pascal \ 2015, \apj, 799, 129. doi:10.1088/0004-637X/799/2/129
\bibitem[Werner et al.(2016)]{Werner16} Werner, G.R., Uzdensky, D.~A., Cerutti, B., Nalewajko, K., \& Begelman, M.~C.\ 2016, \apjl, 816, L8. doi:10.3847/2041-8205/816/1/L8
\bibitem[Petropoulou \& Sironi(2018)]{Petropoulou18} Petropoulou, M. \& Sironi, L.\ 2018, \mnras, 481, 5687. doi:10.1093/mnras/sty2702

\end{thebibliography}
\end{document}